\documentclass[conference]{IEEEtran}
\usepackage[hyphens]{url}
\usepackage{hyperref}
\hypersetup{breaklinks=true, colorlinks,allcolors=blue}
\usepackage[backend=biber,style=ieee,citestyle=numeric-comp]{biblatex}

\IEEEoverridecommandlockouts
\usepackage{amsmath,amssymb,amsfonts}
\usepackage{algorithmic}
\usepackage{graphicx}
\usepackage{textcomp}
\usepackage{xcolor}  
\usepackage{balance}
\usepackage{mathtools}  
\usepackage{orcidlink}  
\usepackage{kantlipsum} 
\usepackage{tabularx}
\usepackage{booktabs}

\def\BibTeX{{\rm B\kern-.05em{\sc i\kern-.025em b}\kern-.08em
    T\kern-.1667em\lower.7ex\hbox{E}\kern-.125emX}}

\begin{document}

\title{Workplace Surveillance and Insider Threat Risk Management: Legal Limits and Privacy Harms}

\author{
Haywood Gelman\,\textsuperscript{†}, 
John D. Hastings\,\textsuperscript{‡}
Suvineetha Herath\,\textsuperscript{‡}, 
and Quentin Covert\,\textsuperscript{§}

  \thanks{\textsuperscript{†}The Beacom College of Computer \& Cyber Sciences, Dakota State University, Madison, SD, USA. Email: haywood.gelman@trojans.dsu.edu}
\thanks{\textsuperscript{‡}The Beacom College of Computer \& Cyber Sciences, Dakota State University, Madison, SD, USA. Email: john.hastings@dsu.edu}
\thanks{\textsuperscript{‡}Department of Information \& Communication Technology, Carl Sandburg College, Galesburg, IL, USA. Email:sherath@sandburg.edu}

\thanks{\textsuperscript{§}The Beacom College of Computer \& Cyber Sciences, Dakota State University, Madison, SD, USA. Email: quentin.covert@dsu.edu}
}

\maketitle
\begin{abstract}
Workplace surveillance is used by organizations to protect corporate assets and monitor employee productivity. This research presents two central arguments on workplace surveillance: although surveillance serves legitimate organizational purposes, over-surveillance can violate legal requirements and data privacy principles; and a primary security objective of workplace surveillance is the detection of insider threats (InT). InT are comprised of individuals with authorized resource access whose intentional or unintentional actions may damage or compromise corporate assets. This paper investigates InT personas to understand behavioral and psychological detection criteria. Employee surveillance tools and techniques are reviewed to characterize the employee surveillance landscape. Workplace privacy laws, examples of over-surveillance, and the resulting privacy harms are addressed. The review identifies research gaps related to over-surveillance, including the generation of excessive alerts that may obscure meaningful InT indicators. The paper concludes with recommendations to improve workplace surveillance transparency, implement InT training programs to improve organizational detection capabilities, and tune InT tools to detect relevant psychological and behavioral indicators.
\end{abstract}


\section{Introduction}

Organizations use workplace surveillance to monitor productivity, protect intellectual property and other sensitive assets, enforce workplace policies, and satisfy legal and regulatory obligations \cite{vatcha_workplace_2020, applin_watching_2013, ghoshray_employer_2013}.
Use of software and hardware tools to monitor activity and network usage are often viewed by employees as an expression of distrust by their employer and a violation of data subject rights \cite{chang_exploring_2017}. 
This view can be tempered in light of electronic surveillance's long history in the workplace \cite{schoenherr_understanding_2020} with collection for a purpose \cite{gdpr_art_5_art_2018}. This history includes physical monitoring with cameras \cite{applin_watching_2013,ball_electronic_2021}, activity sensors \cite{ajunwa_limitless_2017}, digital monitoring of emails \cite{smith_monitoring_2009}, phone calls \cite{pitesa_employee_2012}, log analysis \cite{lyon_toward_2004}, user entity behavior analysis tools (UEBA) \cite{khaliq_role_2020}, and social media monitoring \cite{vatcha_workplace_2020}. Monitoring also extends to personal phones, tablets, computers, and other electronic devices connected to corporate infrastructure \cite{ball_electronic_2021}. Much of the literature is dedicated to the opposition to or limitation of employee surveillance based on privacy principles \cite{ajunwa_limitless_2017,chang_exploring_2017,smith_monitoring_2009,cozzetto_privacy_1997}. This study presents a more nuanced view where focused monitoring is required for organizations to address the true target of employee surveillance policy: insider threat (InT) personas that are responsible for the risk. Perception of organizational risk presented by individuals influence recognition of InT personas, and is directly related to risk of data privacy harms that can occur as a result.

This review examines representative cases in which productivity monitoring, biometric technologies, and the collection of sensitive employee information exceeded their stated purposes or created significant privacy risks \cite{ennis_barclays_2020,devrio_building_2024,hamilton_amazon_2021,sebastian_we_2025,rochford_cothron_2023,hickok_policy_2023}. These cases are analyzed in relation to workplace privacy laws and Solove's taxonomy of privacy harms \cite{solove_taxonomy_2006,herath_privacy_2023} to clarify when legitimate organizational monitoring may become disproportionate or harmful. The review also connects workplace surveillance practices to unintentional, intentional, and opportunity-based insider-threat personas \cite{gelman_toward_2024,schoenherr_multiple_2022,padayachee_framework_2015,maasberg_dark_2020}. It identifies gaps in surveillance transparency, insider-threat education, and the use of behavioral indicators in detection systems, and presents corresponding policy, training, and technical recommendations intended to support more focused and proportionate monitoring. The study is directed by the following research questions: 
\begin{itemize}
    \item \textbf{RQ1}: What types and methods of surveillance do organizations employ to monitor employee electronic and physical activity?
    \item \textbf{RQ2}: What are the legal limits of workplace surveillance and what privacy harms can it cause?
    \item \textbf{RQ3}: What gaps exist in employee surveillance approaches, and how can techniques be refined to reduce privacy harms and unnecessary monitoring while improving outcomes?
\end{itemize}

\section{Methodology}

This study conducted a literature review that assessed employee workplace surveillance tools, practices, limits, and privacy harms. Papers were obtained from database searches and websites in peer-reviewed journals, conferences, books, book sections, documents, theses, law review journals, law review journals, reports, NIST standards documents, statutes, and newspaper articles. Searches were executed with IEEE Xplore, ACM, Google Scholar, and targeted Google searches for relevant newspaper articles and standards documents where appropriate. One preprint was collected but not used. Keywords included ``employee surveillance", ``employee surveillance AI", ``electronic employee surveillance", ``employee monitoring", ``workplace privacy", ``workplace privacy laws", ``workplace privacy state laws", ``employee surveillance tools", and ``privacy harm". Similar search criteria was applied to each source to maintain consistency. Abstracts and overviews for 231 sources were read and relevant, documents compiled, followed by a complete reading of selected sources to attain the 120 utilized in this study, and deduplication applied.
Relevance was determined at two levels: pertinence to the research questions and applicability to the paper's proposals on organizational justification for workplace surveillance. 
While no categorical measurement was implemented for the methodology, layered literature search methods \cite{machi_literature_2022,kumar_research_2019} safeguarded the approach to maintain validity, and to mitigate selection bias through careful search term constructs and organizational discipline \cite{gelman_toward_2024}. 
Table \ref{tab:lit-search} describes the selected databases, records located, records selected, and document types compiled for the study.

\label{lit-search}

\begin{table}[h]
\vspace{-1.2em}
\caption{Search Criteria and Results}
\label{tab:lit-search}
\centering
\footnotesize
\begin{tabularx}{\columnwidth}{
    p{0.14\columnwidth}
    p{0.22\columnwidth}
    X
}
\toprule
\textbf{Database} &
\textbf{Located/Selected} &
\textbf{Document Types} \\
\midrule
NIST                & 3/3  & standards documents \\
ACM                 & 15/7  & conf. paper, journal article \\
IEEE Xplore         & 52/24 & conf. paper, journal article \\
Google Scholar      & 106/54 & conf. paper, journal article, doc., book/book section, thesis, preprint, report \\
website             & 31/16 & law review, news articles, court case \\
statute             & 24/16 & Publications on legislative websites \\
\midrule
\textbf{Total} & 231/120 & 13 \\
\bottomrule
\end{tabularx}
\vspace{-1.0em}
\end{table}

Extensive research on employee monitoring during the COVID-19 pandemic is a singular area of research. Although some examples are provided in the context of employee surveillance, pandemic-era employee monitoring is outside the scope of this paper. General Data Protection Regulation (GDPR) defers workplace privacy laws to its member states \cite{uk_government_countries_2024,gdpr_art_88_art_2018,gdpr_rec_155_recital_2018}. An extensive analysis of workplace surveillance laws in European Union (EU) member states is also out of scope for this paper, although some examples of GDPR fines are included as examples. The Privacy Act of 1974 \cite{us_congress_privacy_1974} addresses aggregation and data management for private citizens' data in federal databases, but is not related to workplace surveillance and is also out of scope. Workplace privacy laws are culturally and regionally relevant, enabling future work for out of scope areas of study.

\section{Literature Review}
\subsection{Insider Threats and Risk Management Overview (RQ1)}

InT risk management addresses causality of behavioral,  psychological, and technological indicators through educational perspectives \cite{gelman_toward_2024}. It also describes InTs as individuals with sanctioned access to information, systems, and infrastructure \cite{padayachee_conceptual_2013,gelman_toward_2024} who compromise, damage, or steal intellectual property, confidential, or proprietary data. Research also defines common personas that determine an individual's propensity toward sub-criminal behavior \cite{harms_exposing_2022}. Understanding personas is based on classification of characteristics, motivations, and behaviors (CMB) that drive insiders to commit InT acts \cite{mills_current_2018}. CMB include policy ignorance and computer misuse \cite{warkentin_behavioral_2009,willison_beyond_2013}, financial and revenge motivations \cite{willison_disentangling_2018,padayachee_exploratory_2022}, and Dark Triad traits \cite{harms_exposing_2022,maasberg_dark_2015}. Personas include unintentional insider threat (UIT), intentional insider threat (IIT) \cite{schoenherr_multiple_2022}, and insider threat of opportunity (ITO) \cite{gelman_toward_2024,padayachee_framework_2015,maasberg_dark_2020}. As a result of this combination of factors, and despite the decline in malicious InTs in 2025 and the lower likelihood overall of insider activity \cite{verizon_dbir_2026_verizon_2026}, InTs possess influence that far outweighs their small numbers due to their trusted status \cite{gelman_toward_2024}. Greitzer et al. \cite{greitzer_identifying_2012} determines a list of ``psychosocial indicators''  describing patterns of behavior demonstrating precursors of InT risk. Previously described employee surveillance tools collect information on physical movement and digital monitoring. In addition, overarching crime and prevention theories are proposed, including situational crime theory \cite{mekonnen_privacy_2015}, cognitive dissonance \cite{padayachee_insider_2015}, general detection and social bond theories \cite{mills_current_2018}, and protection motivation theory \cite{colwill_human_2009} define the framework around which seminal authors describe the InT risk management landscape. This descriptive research of diverse InT factors proposes that the true target of employee surveillance is the prospect of InTs within an organization, and should not be construed as a quantitative contrast. A brief description of the three most common InT personas follows to position InTs within the structure of employee surveillance.

\subsubsection{Unintentional InT}

Schoenherr \cite{schoenherr_multiple_2022} attributes insider threats to ``individual motivation and social context.'' UIT are characterized as possessing an impaired intent to commit an act, while lacking requisite training to prevent one \cite{greitzer_analysis_2014}. Characteristics also include difficulty in following policy, and a propensity to overshare information online due to cognitive perceptions of stress and time \cite{schoenherr_multiple_2022}. UIT make up the largest percentage of InT, with 80\% of UIT actions attributable to human error, and to a lesser degree, inattentiveness and fatigue \cite{cert_insider_threat_team_unintentional_2013}. Employee surveillance for indicators of UIT include ``reporting culture'' \cite{khan_understanding_2022}, internet monitoring \cite{colwill_human_2009}, policy adherence and productivity monitoring \cite{greitzer_identifying_2010}.

\subsubsection{Intentional InT} 

IIT are seen in much smaller numbers than UIT, but have the ability to inflict more harm. This is due to differing CMB \cite{mills_current_2018} and social paths \cite{schoenherr_multiple_2022}. CMB include a disregard for policies \cite{greitzer_identifying_2012}, feelings of resentment toward the organization \cite{dtex_systems_int_2024}, financial hardship \cite{schoenherr_multiple_2022}, and attempts to gain access to content in excess of given permissions \cite{alawneh_defining_2011}. Per DTEX \cite{dtex_systems_int_2024}, 15\% of separating employees stole sensitive intellectual property when they departed the organization. Employee surveillance for indicators of IIT include log analysis \cite{lyon_toward_2004,brown_predicting_2013}, social network monitoring \cite{skaik_using_2021}, human relationship monitoring \cite{huertas-baker_exploring_2022}, machine learning \cite{brdiczka_proactive_2012}, and UEBA tools \cite{khaliq_role_2020}.

\subsubsection{InT of Opportunity}

Though the ITO persona was identified by penumbras in the literature \cite{gelman_toward_2024}, it was first introduced by Padayachee \cite{padayachee_framework_2015} as a malicious insider who waits for the opportunity to commit an InT act. Aspects of the ITO persona were also discussed in related work \cite{maasberg_dark_2015,maasberg_dark_2020,harms_exposing_2022} as Dark Triad traits of narcissism (self-importance), Machiavellianism (arrogance and manipulation), and psychopathy (lack of self control). Characteristics of ITO include heightened intelligence, calm demeanor, and the observational and technical skills to observe an opportunity to act, similar to Dark Traits \cite{maasberg_dark_2020}. Padayachee's study on minimization of opportunity-based InT \cite{padayachee_framework_2015} drew connections to the fraud triangle stating the unique nature of opportunity-driven InT activity along with Cressey's pressure (purpose for committing a crime) and rationalization (justification) \cite{cressey_other_1953}.

ITO are difficult to detect due to these characteristics, but can be detected through social engineering \cite{xiangyu_social_2017,cert_insider_threat_team_unintentional_2014}, UEBA tools \cite{khaliq_role_2020}, digital semantic analysis \cite{brown_predicting_2013}, deep log analysis \cite{lyon_toward_2004}, and behavior modeling \cite{kim_insider_2019}. Perhaps the most unique of these risks is social engineering, as ITO uses social engineering skills not to influence an individual outside the company to compromise private data or steal credentials, but to influence an internal UIT or IIT to commit an InT act \cite{sum_its_2025,sharma_analyzing_2024,xiangyu_social_2017,uebelacker_social_2014}. InT personas as described demonstrate surveillance tool intersection between employees and insider threats.

\begin{table*}[ht]
\caption{Connecting InT, Surveillance Intent, Tools, Laws, and Privacy Harms}
\label{tab:int-connection}
\centering
\footnotesize
\begin{tabularx}{\textwidth}{
    p{0.21\textwidth}
    p{0.10\textwidth}
    p{0.16\textwidth}
    p{0.23\textwidth}
    p{0.17\textwidth}
}
\toprule
\textbf{Surveillance Intent} &
\textbf{InT Persona} &
\textbf{Tools} &
\textbf{Laws/Legal Restraint} &
\textbf{Privacy Harms} \\
\midrule
Personal account monitoring & IIT, ITO & Account surveillance & IRPWA \cite{legislative_is_irpwa_2024} NYS-201-I \cite{nys_senate_nys_2024} NYS-C52 \cite{ny_state_senate_nys_2022}(personal acct. access)& surveillance, intrusion\\
Theft and productivity detection & IIT, ITO & Audio/video mon. & MESA \cite{maine_legislature_524_2026}, NYS-C52 (advance notice) & surveillance, aggregation, identification, exposure, intrusion\\
Perimeter and physical security & IIT, ITO & Camera surveillance & NYS-203-C, MESA (prohibited areas) & surveillance, identification\\
Banned sites, unsanctioned sites & UIT, IIT & Inter/intranet monitoring & NYS-52C, MESA, (advance notice) & surveillance, identification\\
Unauthorized access & UIT, IIT & Keyloggers, location tracking & MESA (locator software) & surveillance, interrogation, identification\\
Analyze access intent & IIT, ITO & Deep log analytics & MESA \cite{maine_legislature_524_2026}, NYS-C52 (advance notice) & surveillance, aggregation, identification, exposure, intrusion \\
Analyze behavioral intent & IIT, ITO & UEBA & MESA \cite{maine_legislature_524_2026}, NYS-C52 (advance notice) & surveillance, aggregation, identification, exposure, intrusion\\
Analyze access patterns & IIT, ITO & Biometric & BIPA \cite{legislative_is_bipa_2024} & surveillance, identification, secondary use \\
\bottomrule
\end{tabularx}
\vspace{-2.2em}
\end{table*}

Table \ref{tab:int-connection} synthesizes the connections between surveillance intent, InT personas, the tools deployed to perform surveillance, the laws intended to govern their use, and the privacy harms that result from over-surveillance. The table describes legal gaps where numerous activities are ungoverned. Employers need to proportionally apply surveillance tools and techniques based on the observed risk to the organization, collecting only what is required to perform the task, while maintaining compliance with laws \cite{macnish_ethics_2026}.

\subsection{Employee Surveillance Tools and Practices (RQ1)}

Several categories of employee surveillance tools collect, analyze, and disseminate data for different purposes. Email surveillance is designed to prevent the inappropriate use of corporate email to maintain outside-work activities, protect against harassment, and prevent phishing attacks \cite{vatcha_workplace_2020,ball_electronic_2021,thompson_workplace_2023}. AI-driven audio and video monitoring, as well as electronic communications, enables employers to draw machine-controlled insights from large datasets \cite{corvite_data_2023,hickok_policy_2023}. Camera-based monitoring is intended to capture productivity data locally on personal computers and in open areas to ensure productivity and policy adherence \cite{g_real-time_2023}. Internet monitoring, web filtering, and cloud access security brokers are designed to observe and prevent access to unsanctioned websites, file sharing sites, banned content, and excessive use \cite{posey_employees_2024,nazarov_intelligent_2023}. During increased use of surveillance technologies during the COVID-19 pandemic, employers used advanced internet monitoring software in addition to keyloggers as a way of tracking potentially damaging actions on the part of an unintentional lapse of judgment or awareness \cite{charbonneau_empirical_2020,thompson_workplace_2023}. Deep log analytics are implemented through machine learning techniques to discover hidden intent through behavioral system access techniques with security incident event management (SIEM) and syslog tools \cite{kim_insider_2019,bin_sarhan_insider_2023,gopal_analysis_2022}. UEBA tools are used in conjunction with deep log analytics to create a more complete InT attack visualization \cite{khaliq_role_2020}. Last, biometric data collection and algorithmic decision making tools are used by employers to validate employability and exclude applicants from consideration based on AI data analysis \cite{hickok_policy_2023,kellogg_algorithms_2020}.

\subsection{Surveillance Limits (RQ2)}
As previously described, the three primary purposes of employee surveillance are to monitor employee productivity, mitigate electronic abuse of corporate resources, and to detect InT risk to information, systems, and infrastructure. Surveillance limits are guided by their adherence to enacted laws for the purpose of accountability, detailed in the section that follows. In 2020, Barclays risked a maximum \$1.2B USD UK-GDPR fine for implementing surveillance software to monitor the amount of time employees remained working at their desks throughout the day \cite{ennis_barclays_2020}. When originally implemented, the tool provided anonymized tracking to leadership, functionality that was eventually augmented to allow tracking of individual productivity \cite{ennis_barclays_2020}. In a similar example, pre-GDPR, Barclays installed heat and motion sensors under the desks of office workers to monitor the amount of time they spent physically at their desks \cite{ball_electronic_2021}.

Amazon used web cameras to track employees against a metric measuring time away from their desks \cite{staff_update_2021}, as well as using AI cameras to track safety of delivery drivers and punished them for violations as determined algorithmically \cite{devrio_building_2024}. Amazon France Logistique was fined €32 million by National Commission on Informatics and Liberty (CNIL) in 2023 for monitoring warehouse employees through the use of handheld scanners \cite{sebastian_we_2025}. The scanner system itself was not designed as a surveillance tool, but was repurposed by collecting scans-per-second metrics for each scanner \cite{ap_news_france_2024}. The White Castle restaurant chain was successfully sued by an employee claiming violation of biometric rights by the chain in how the company collected and used fingerprint data related to paychecks in Illinois \cite{rochford_cothron_2023}. 

In 2020, the H\&M retail clothing chain based in Germany maintained excessive surveillance on employees in Nuremberg that included personal details on health, time off, and family problems \cite{bbc_hm_fine_hm_2020}. Accidentally exposed data on a network drive maintained by management enabled full visibility of its contents and revealed the extent of the surveillance \cite{lensdorf_hm_2020}. H\&M was fined €35.3 million USD by the German EU Data Protection Authority \cite{lensdorf_hm_2020}. In these examples of surveillance limits, most were resolved by the organization changing surveillance direction, but examples include a potential GDPR fine \cite{ennis_barclays_2020} and an actual fine \cite{lensdorf_hm_2020}.

\subsection{Employee Workplace Privacy Laws (RQ2)}

\subsubsection{Workplace Privacy Laws Overview}

U.S. state data privacy laws are described universally as a patchwork of state laws with no comprehensive law governing data privacy \cite{cozzens_patchwork_2022}. The same can be stated regarding workplace privacy laws in the U.S, which according to \cite{hirsch_j_m_university_2020}, ``These privacy laws mirror the shortcoming of our workplace regulatory
system as a whole: a patchwork of federal and state legislation that leaves huge
gaps in protection for a large swath of workers under numerous legal situations.'' The differences in data privacy laws creates confusion, but regional differences in laws can at least establish a baseline of compliance \cite{cozzens_patchwork_2022}. The workplace privacy laws enacted by 28 states are organized into key thematic areas. Legal themes include personal social media and online account protection, employee and applicant personal data protection, employer digital monitoring notice requirement, biometric data, and personal vehicle tracking notification. 

\subsubsection{Consent to Recording}
Every state in the U.S. has enacted laws that govern one, two, or all party consent to either listen/observe a phone conversation unknown to the other party (eavesdrop), record a conversation, or both \cite{orourke_electronic_1998,mathiessen_laws_2024}. The Electronics Communications Privacy Act (ECPA) also prevents wiretapping at the federal level for one-party consent \cite{us_congress_ecpa_18_1986}. Arguments have been made as to the constitutionality of preventing listening/recording under the First Amendment, but legal precedent has positioned one and two party consent laws as outside the sphere of first amendment litigation, in contrast to some legal scholars' opinions \cite{kaminski_privacy_2017}. The legal requirement states that one or both parties to a conversation, state-dependent, must consent to recording for the recording to be legally permissible \cite{mathiessen_laws_2024}. In the case of a customer support organization informing a customer that their conversation may be recorded for training, the customer can simply state they do not wish the recording to take place. In the case of clandestine recording without consent, the party recording the conversation would be in legal peril, in some states, as a felony \cite{justia_nh__570-a2i-a_2023_2024}. Consent to recording is an underlying factor in the state workplace privacy laws that follow. The privacy laws discussion is restricted to laws enacted in states specific to workplace privacy, excluding states like Massachusetts and Vermont that have provisions for workplace privacy nested in other laws. For example, Vermont does not have a consent to record law but defers to federal law \cite{mathiessen_laws_2024}.

\subsubsection{Maine Employer Surveillance Act (MESA)}

Enacted in 2026, this law specifies privacy rights for employees in the workplace. It designates notification requirements when monitoring occurs, the monitoring type, limitations on surveillance in the employee's home, the type of data collected, and a specified purpose \cite{maine_legislature_524_2026}. Specifically, the law excludes protections in home-care settings where employee monitoring is required by the service \cite{524_summ_sen_2026}. The law also enforces fines on employers for non-compliance, allows employees the right to refuse access to personal accounts for surveillance purposes, and requires employers to disclose the existence of surveillance practices during interviews for a role \cite{maine_legislature_524_2026}.

\subsubsection{Illinois Right to Privacy in the Workplace Act (IRPWA)}

IRPWA \cite{legislative_is_irpwa_2024} was the first comprehensive workplace privacy law enacted in the U.S. Enacted in 1992, the law focused on specific allowed and prohibited actions by employers, and is similar to MESA. Like MESA, the law does not prohibit an employer from surveilling employees using corporate assets, however, the law prohibits employers from requiring access to employees and potential employees' personal electronic accounts through open and surreptitious means \cite{legislative_is_irpwa_2024} or placing inquiries regarding workers compensation claims \cite{ford_employee_2016}. The law prohibits employers from retaliation against employee refusal to comply with access requests, enables employers to monitor employees' electronic use of corporate resources, and requires employers to maintain well-documented policies on surveillance practices \cite{legislative_is_irpwa_2024}. The New York State workplace privacy laws that follow contain a similar personal account access provision, similar to IRPWA. 

\subsubsection{NY State Workplace Privacy Laws}

Unlike MESA and IRPWA that enumerate rules in a single law, New York State's workplace privacy laws are organized under labor and civil rights laws by the right to privacy. New York State Labor Law Chapter 31, Article 7, Section 201-I  \cite{nys_senate_nys_2024} describes the ``request for access to personal accounts", similar to IRPWA. Provisions include employee protections from providing access to personal devices or content, while maintaining an employer's ability to electronically monitor devices that utilize corporate resources. New York State Labor Law Ch. 31, Art. 7, Sect. 203-C \cite{ny_state_senate_nys_2014} describes ``employee privacy protection''  that prohibits video recordings in locations with an expectation of privacy. Last, New York State Civil Rights (CVR) Ch. 6, Art. 5, Sec. 52-C*2 describes ``Employers engaged in electronic monitoring; prior notice required". Provisions of this law require employers engaging in workplace surveillance to monitor electronic or phone transmissions of any kind provide written notice to new and existing employees on the nature and purpose of the surveillance, with noted exceptions similar to MESA and IRPWA \cite{ny_state_senate_nys_2022}. In the example of a New York Southern District case, ``Pure Power Boot Camp, Inc. v. Warrior Fitness Boot Camp, LLC" \cite{PurePower2008}, the court found that accessing an employee's personal email accounts without authorization violated the Federal Stored Communications Act and prevented the use of the improperly obtained emails in the litigation. 

\subsection{Privacy Harms (RQ2)}

Privacy harms may arise when privacy rights, laws, or principles are violated, resulting in physical, financial, reputational, emotional, or other harm to an individual. The concept of risk related to privacy harms is considered difficult to prove as actual harm \cite{solove_risk_2016,herath_privacy_2023}, although several privacy harms described in previous over-surveillance examples describe actual harm arising from violation of a workplace privacy law. Most state workplace privacy laws reserve legal action for state agencies, while private right of action and surveillance notification to employees is granted in Delaware, New York, Maine, Connecticut, and Illinois \cite{de_legislature_delaware_2025,ny_state_senate_nys_2022,maine_legislature_524_2026,ct_legislature_chapter_2024,legislative_is_irpwa_2024}. 

\subsubsection{Information Collection Activity}

Surveillance and interrogation are harms born of surreptitious observation and rigorous questioning regarding private information \cite{solove_taxonomy_2006}. New York State Civil Rights (CVR) Ch. 6, Art. 5, Sec. 52-C*2 and MESA require employers to notify employees regarding the nature, extent, and purpose of surveillance \cite{ny_state_senate_nys_2022,maine_legislature_524_2026}. MESA, IRPWA, and NYS 201-I contain specific rules prohibiting employers from requesting access to personal online accounts, establishing a baseline for adherence \cite{maine_legislature_524_2026,legislative_is_irpwa_2024,nys_senate_nys_2024}. Barclays, White Castle, and Amazon caused a surveillance privacy harm in two ways: collection for a purpose and secondary use \cite{hamilton_amazon_2021,rochford_cothron_2023}. H\&M violated surveillance harm when they collected personal information from employees and stored for use by management in decision making \cite{lensdorf_hm_2020}. The company also violated interrogation harm by engaging in conversations to gather more information for employment purposes \cite{bbc_hm_fine_hm_2020}.

\subsubsection{Information Processing Activity}

Information processing activities are harms caused by improper use of data post-operation that exposes the data subject to harm \cite{solove_taxonomy_2006}. Aggregation, identification, and secondary use are particularly problematic in this set of activities, as data under these harms can be used individually or collectively to 
cause actual harm \cite{solove_new_2008}. H\&M caused an aggregation harm by collecting personal data on employees and storing the data on a server only accessible to management \cite{lensdorf_hm_2020}. Barclays and Amazon caused an identification harm by using hardware and software to monitor employee productivity in a way that could identify individual employees \cite{hamilton_amazon_2021,ennis_barclays_2020}. This ``downstream harm'' \cite{solove_taxonomy_2006} occurs when a primary harm (such as aggregation and identification) causes another harm. 

In the case of Amazon \cite{ap_news_france_2024}, this caused a downstream harm of secondary use, a harm created by now-identified data for another, previously unstated purpose. Amazon's use of handheld scanners to track productivity had a secondary use in employment decisions. The Illinois Legislature recently repealed Sections 12 and 13 of IRPWA governing the use of the E-Verify system in employment decisions over concern for discriminatory use \cite{legislative_is_irpwa_2024}. New York City Local Law 144 of 2021 \cite{nycc_local_law_144_of_2021_new_2021} was enacted specifically to address the use of Automated Decision Systems (ADS) in employment decisions, which extends to employee surveillance. The key issue in both state laws is the concern of bias and discrimination in employment decisions based on personal information gathered by the systems. Related work \cite{kellogg_algorithms_2020,devrio_building_2024,hickok_policy_2023} concurs while describing the need for accuracy and accountability in AI-based ADS to avoid ``algorithmic harm'' \cite{devrio_building_2024}. The concern also exists in potential for violation of Title VII of the Civil Rights Act \cite{wilson_building_2021}, Americans with Disabilities Act \cite{timmons_pre-employment_2022}, Age Discrimination in Employment Act \cite{scherer_applying_2019}, the Pregnancy Discrimination Act \cite{schwarcz_health-based_2021}, and the Genetic Information Nondiscrimination Act \cite{rothstein_predictive_2020} in use of ADS and AI-based decision systems.

\subsubsection{Information Dissemination Activity}

Information dissemination is defined by Solove \cite{solove_taxonomy_2006} as ``revelation of personal data or the threat of spreading information''. Harms arise when private data is promulgated in a manner that causes physical, reputational, or financial loss. The H\&M enforcement action caused the harm of exposure by revealing private data on a server to all internal employees \cite{lensdorf_hm_2020}.

\subsubsection{Invasion Activity}

As a privacy harm, intrusion arises when workplace monitoring or employer demands extend beyond normal work hours. Disruptions can occur outside of work hours or when an employer encroaches on an employee's personal time that is not part of a known work expectation \cite{solove_taxonomy_2006}. No state workplace privacy laws explicitly define work hours, but analogues exist. \cite{maine_legislature_524_2026} describes ``audiovisual
monitoring in an employee's residence or personal vehicle or on the employee's property'' as a workplace surveillance restriction only if required for work, as in a home care worker. The Connecticut electronic surveillance law refers to employee-managed premises only \cite{ct_legislature_chapter_2024}, but other non employee-surveillance laws may be applicable in this area.

The literature argues for reducing employee surveillance to what is necessary to minimize collection noise in the academic sense \cite{ciocchetti_eavesdropping_2011}. As noted previously, employee surveillance is performed to assess employee productivity, to ensure adherence to policy that protects systems, information, and infrastructure, and to mitigate insider threats. Information Technology (IT) departments collect an abundance of log data from SIEM, syslog, and UEBA that often creates log and alert fatigue from alerts with minimal value \cite{baruwal_chhetri_towards_2024}, a situation where important alerts can be missed due to the sheer volume of log data \cite{ahmed_reduction_2021,vaarandi_how_2022}. Excessive employee monitoring can lead to violation of laws, clarification of policies, violation of data privacy principles and harms, loss of reputation, and loss of employee trust. As previously noted, MESA and New York State employee surveillance laws require employers to provide notice to employees as to the nature, extent, and purpose of monitoring. For these reasons, the employee surveillance awareness practice exhibited in MESA and NYS laws creates a more cyber-aware environment, and may itself reduce the likelihood and severity of UIT \cite{colwill_human_2009,darcy_user_2009} and force IIT and ITO to use other means of accomplishing their InT goals \cite{pitesa_employee_2012}. In so doing, excessive logging may be reduced to enable focused analysis on insider threats in organizations, work that could be supported by future empirical research. 

\section{Research Gaps (RQ3)}

\subsection{Inconsistent Workplace Surveillance Program Visibility}

Awareness of InT programs by employees may reduce the likelihood and severity of attack by redirecting the most sinister InT, and forcing communication methods to outside channels \cite{pitesa_employee_2012}. Outside channels may include flash drives, flash card readers, cloud storage, VPN to unsanctioned destinations, and unsanctioned file share locations. Excessive logging creates alert fatigue for operators \cite{alzaabi_review_2024}, but following data privacy principles may lead to reduced logging and greater operator effectiveness \cite{gdpr_art_5_art_2018}. Visibility and accountability in surveillance may provide employees with greater visibility into surveillance practices, but it should also increase awareness of potential risks to employees and organizations from insider threats. New York State \cite{ny_state_senate_nys_2022}, Connecticut \cite{ct_legislature_chapter_2024}, and Delaware \cite{de_legislature_delaware_2025} have explicit laws that require employers to notify employees on the existence and extent of surveillance programs. New Jersey law \cite{nj_legislature_nj_2022} relates to vehicle tracking, and Illinois' Biometric Information Privacy Act (BIPA) \cite{legislative_is_bipa_2024} is specific to biometric monitoring. Although the California Consumer Privacy Act (CCPA) \cite{california_legislative_information_california_2020} requires notice to consumers of data collection, the law does not require notice to employees by employers. This gap can greatly benefit from empirical research.

\subsection{Gaps in Insider Threat Education Programs}

This paper described education as a critical factor both in reducing the number of InT incidents and improving mitigation. The concern is in how InT personas are recognized, and how knowledge of InT is shared through organizations \cite{gelman_toward_2024}. Few comprehensive security education training and awareness (SETA) programs exist for InT training, most of which apply to federal employees \cite{gelman_toward_2024}. Int training programs must consist of ontology, framework, persona, policy and methods to ensure success and assist organizations in mitigating insider threats while minimizing surveillance \cite{cdse_int_training_insider_2023}.

\subsection{Logging Systems Lack High Fidelity Behavioral Indicators}

Reducing log noise through awareness of workplace surveillance programs may reduce the number of unnecessary logs managed by IT departments, but logging systems require improved sensitivity for behavioral indicators of compromise (BIoC) to increase fidelity and mitigate psychological and behavioral threats \cite{bin_sarhan_insider_2023}. This process requires integration of UEBA logs with SIEMs and behavioral threat intelligence feeds to correlate InT incident behavior \cite{khaliq_role_2020}.

\section{Discussion: Proportionality (RQ3)} 

Insights imply that workplace surveillance oversight should be performed at three levels. Proportionality in data collection and awareness of surveillance through legal notice demonstrate the need for targeted collection. Insider threats delineate the factors that place organizations internally at risk and establish learning opportunities to limit data collection. Data collection methods describe the technical aspects of collection, where tool calibration can reduce collection to what is necessary. 

\subsection{Legislate State Workplace Surveillance Laws}

Most state workplace surveillance laws on do not require notification, creating confusion among employees as to their rights.  In support of this, comprehensive workplace privacy laws similar to MESA should be passed at the state level to recognize regional workplace privacy differences. A comprehensive federal workplace privacy law would minimize policy confusion, but may not account for regional perceptive differences regarding workplace privacy. For example, 20 U.S. states have codified sweeping data privacy laws where 4 are focused on protections for specific data \cite{iapp_priv_legis_tracker_us_nodate}. Data privacy laws can be used as a legislative framework for workplace surveillance laws, modified for regional significance. A gap also exists in establishing geographical workplace surveillance limits, and defining normal work hours. Data privacy principles from CCPA \cite{california_legislative_information_california_2020} and GDPR Art. 5 \cite{gdpr_art_5_art_2018} can be used as a guide for legislatures when crafting laws to define the limits of workplace surveillance.

\subsection{Develop Insider Threat Education Platform}

Most cyber awareness programs exclaim a need for InT training on InT identification through psychological and behavioral means first, technical second \cite{gelman_toward_2024}. Ontologies define the lexicon of InT \cite{padayachee_taxonomy_2012} while frameworks determine detection methodologies \cite{ikany_symptomatic_2019,nurse_understanding_2014}. Personas tune frameworks to define tactics and techniques of detection \cite{schoenherr_understanding_2020,padayachee_exploratory_2022,maasberg_dark_2020,harms_exposing_2022} while policy drives the application of training \cite{darcy_user_2009,ho_behavioral_2009} to provide a complete training program \cite{gelman_toward_2024,bloom_taxonomy_1956,wilson_building_2003,cappelli_cert_2012,cdse_int_training_insider_2023,petersen_workforce_2020}. InT training programs, therefore, enable people across an organization to perform the role of InT sensors \cite{greitzer_modeling_2011}. Training programs also serve to inform the community on laws in place to prevent privacy harm.

\subsection{Improving InT Detection with BIoCs}

Although not researched empirically in this literature review, existing InT tools should be human-evaluated and aligned with organizational objectives and legal obligations. BIoCs should be tuned to data privacy principles \cite{gdpr_art_5_art_2018}, limited to work-related behaviors, and exclude personality traits, behavioral traits unrelated to technical indicators, protected status, and personal activities. With these restrictions, InT tools should be evaluated for tuning to ingest psychological and behavioral IoCs to reduce false positives and mitigate privacy harm. Tuning indicators will enable UEBA and SIEM tools to distinguish InT from non-InT indicators to improve true-positive match results while adhering to ethical data collection. This could be investigated through machine learning techniques to analyze anomalies for distinction from baseline behavior \cite{padayachee_framework_2015,bin_sarhan_insider_2023,alzaabi_review_2024}. Padayachee \cite{padayachee_framework_2015} stated the need for mitigating InT activity through organizational behaviors, reducing emotional triggers often associated with InT activity, and improving the effectiveness of SETA programs. 

\section{Conclusion}

This paper examined the impact and relevance of workplace surveillance. The literature review identified the three main drivers for organizations to conduct workplace surveillance: policy adherence, protection of intellectual property, and protection against insider threats. The paper discussed InT personas, and connected them to surveillance methods. The paper also described examples of over-surveillance practices, deliberated laws enacted to prevent violation of workplace privacy, and addressed privacy harms mapped to examples. Research gaps in the literature were identified to improve U.S. workplace privacy visibility, advise on implementation of an InT-based educational program for organizations, and recommend ways to improve InT monitoring tools through reduction of log volume, supported by empirical research. This review argues the merits of transparency in workplace surveillance, limiting surveillance data collection to only what's necessary to address documentable organizational risks, and to clearly define the legal limits of employer workplace surveillance practices. 

\printbibliography

\end{document}